\documentclass{article}

\usepackage{graphicx}
\usepackage{cite}

\newcommand{\lesssim}{\,\raisebox{-1mm}{$\stackrel{{\textstyle < }}
                                           {{\textstyle \sim}}$}\,}
\newcommand{\gtrsim}{\,\raisebox{-1mm}{$\stackrel{{\textstyle > }}
                                           {{\textstyle \sim}}$}\,}

\begin{document}
\title{Searching for Extra Dimensions and New String-Inspired Forces
in the Casimir Regime$^{*}$ }

\author{Dennis E. Krause$^{1,2}$ and Ephraim Fischbach$^{2}$ \\
\,\\
$^{1}$Wabash College, Crawfordsville, IN 47933-0352 \\
$^{2}$Purdue University, West Lafayette, IN 47907-1396}

\maketitle

\begin{abstract}
The appearance of new fundamental forces and extra-dimensional modifications
to gravity in extensions of the Standard Model
has motivated considerable interest in testing Newtonian gravity at short
distances ({$\lesssim 10^{-3}$} m).  Presently a number of new gravity
experiments are searching for non-Newtonian effects in the ranges
$\sim 10^{-4}$--$10^{-3}$ m.  However, as challenging as these experiments
are, formidable new obstacles await the next generation of experiments which
will probe gravity at distances
$\lesssim 10^{-4}$ m where Casimir/van der Waals forces become dominant. 
Here we will review the motivation for conducting such very short distance
gravity experiments, and discuss some of the new problems that may arise in
future experiments. Finally, we suggest schematic designs for null
experiments which would address some of these problems using the
``iso-electronic'' and ``finite-size'' effects. 
\footnotetext[1]{To appear in {\em Testing General Relativity in Space:
Gyroscopes, Clocks, and Interferometers,} edited by C. L\"{a}mmerzahl, C.
W. F. Everitt, F. W. Hehl (Springer-Verlag, 2000).}

\end{abstract}

\section{Introduction}
	When Isaac Newton formulated his law of universal gravity over 300 years
ago, he provided the first mathematical description of one of the fundamental
forces of nature.  Yet, physicists have realized only relatively recently
that tests of Newtonian gravity can still provide a unique window into new
physics  
\cite{Fischbachbook,Ciufolini,Will,Chen,Bibliography,AdelReview,Fujii}.   
Within the past 20 years, experimentalists have put
Newtonian gravity to the test for distance scales $10^{-3}$--$10^{15}$ m by
searching for violations of the weak equivalence principle (WEP) and inverse
square law (ISL).  The fact that no such violations have been observed
places stringent constraints on extensions of the Standard Model that
would naturally lead to such effects \cite{Fischbachbook}.   Despite this
effort, a number of authors have pointed out that very little is known
of the validity of Newtonian gravity at distances $\lesssim 10^{-3}$ m
\cite{Price,Long,Onofrio,Beane,Giudice}.  Several experimental groups are currently
attempting to extend these limits down to $10^{-4}$
m \cite{Long,Carugno,Deufel}, which is near the point where Casimir/van der
Waals forces overcome gravity to become the dominant force between neutral,
non-magnetic bodies.  This strong intermolecular force background will become
a major challenge for experimentalists who attempt to probe gravity at much
smaller distances.  The purpose of this paper is show that such experiments
are worth the effort despite the new difficulties, and to suggest ideas
which may be useful in detecting new forces of gravitational strength
against a strong intermolecular force background.

	  We begin by examining the theoretical motivation for studying
Newtonian gravity and the phenomenology used to characterize non-Newtonian
effects which would be the signal of new physics in a gravity experiment. 
After briefly reviewing the current constraints on new forces achieved by
longer distance gravity experiments ($\gtrsim 10^{-3}$ m), we will see that
new problems arise when one attempts to set comparable limits in the
Casimir/van der Waals regime ($\lesssim 10^{-4}$~m) where intermolecular
forces become large.  We then investigate these problems quantitatively by
computing the forces between two parallel plates over a range of
separations.  Finally, we propose two schematic designs for null experiments
designed to subtract out the unwanted intermolecular and gravitational
backgrounds using the ``iso-electronic'' and ``finite-size'' effects. These
will hopefully allow one to search for signals of new physics at very short
distances in the presence of Casimir/van der Waals forces.

\section{Theoretical Motivation and Phenomenology}

\subsection{Overview}

	The Standard Model currently provides an adequate
description of the electromagnetic, weak, and strong interactions within the
framework of quantum field theory.  However, a consistent description of
quantum gravity has yet to be formulated despite intense work over the past
fifty years.  The lack of a quantum theory of gravity currently provides
much of the motivation for studying extensions of the Standard Model which
would bring all the fundamental forces into the quantum realm.  In fact, many
believe that the present Standard Model is really only an effective theory
which would be superseded at much higher energies by a more fundamental
theory, such as string or M-theory
\cite{Polchinski,Kaku}.  One of the main problems with these more
fundamental theories is that, despite their purported mathematical beauty,
many of their principal consequences lay far beyond the reach of most
foreseeable experiments.  It is therefore vital to investigate the
low-energy limits of these fundamental theories to allow experimentalists
the opportunity to constrain the proliferation of models which would
otherwise go unchecked. 

			It is against this backdrop that one should view recent and future
experiments testing Newtonian gravity.  Many extensions of the Standard
Model, including string theory, contain new light bosons which would manifest
themselves as new fundamental
forces\cite{Fischbachbook,AdelReview,Fujii,Giudice}.  These new forces would
compete with the other known forces, but they would  most likely be revealed in
a gravity experiment for several reasons.  First, in many ways gravity remains
the least understood of the fundamental forces and is relatively untested over
a wide range of distance scales.  Second, any new forces probably couple very
weakly with matter---otherwise they would have been seen already.  Since
gravity is by far the weakest fundamental force, it sets a natural scale from
which to measure new weak forces.  Third, there are two signatures of gravity
which help one extract a signal from the background of other forces: 1) Since
the  gravitational force couples to mass, it obeys the weak equivalence
principle (WEP), so violations of the WEP would indicate the presence of a
non-gravitational force.  2) The Newtonian gravitational force between point
particles obeys an inverse square law (ISL), hence any departures from the ISL
might be attributed to new forces.  Finally, Newtonian gravity is the
weak-field, non-relativistic limit of General Relativity, a theory in which
gravity is seen as a  manifestation of spacetime that has been curved by
mass-energy.  Therefore, any theory that impacts our understanding of space and
time must involve gravity.  This is important because string theory requires
that there exist more than three spatial dimensions, the new extra dimensions
being rendered invisible to current experiments by some yet-to-be-understood
mechanism.  As will be discussed below, recent models suggest that these new
dimensions may modify Newtonian gravity at short, but macroscopic, distances.

\subsection{Yukawa Potentials}

		The form of the violations of Newton's law of gravity arising from new
physics will be to some extent model dependent, but one finds in practice
that most theories yield modifications that have similar generic features
\cite{Fischbachbook}.  For example,
suppose there exists a new vector field $A^{\mu}(x)$ which couples to
fermions via the Lagrangian density
\begin{equation}
{\cal L}(x) = if\overline{\psi}(x)\gamma_{\mu}\psi(x)A^{\mu}(x).
\end{equation}
Here $f$ is the dimensionless vector-fermion coupling constant ($\hbar = c =
1$) and $\psi(x)$ is the fermion field operator.  If two fermions 1 and 2
exchange a single vector boson with mass $m$  via this coupling,
the lowest order interaction in the non-relativistic limit yields a Yukawa
potential
\begin{equation}
V_{v}(r) = \pm\frac{f_{1}f_{2}}{4\pi}\frac{e^{-mr}}{r},
\end{equation}
where ``$+$'' (``$-$'') indicates that the force is repulsive (attractive)
between like charges.  If this was electromagnetism, massless photons give
$m = 0$, and for electrons $f = -e$.  In the units we use, the range of the
interaction is $\lambda \equiv 1/m$, so that $m \sim
10^{-5}$~eV gives $\lambda \sim 1$~cm, for example.  If the exchanged bosons
were scalars instead of vectors, one arrives at an attractive Yukawa
potential between identical fermions:
\begin{equation}
V_{s}(r) = -\frac{f_{1}f_{2}}{4\pi}\frac{e^{-mr}}{r},
\end{equation}
where $f_{i}$ is now the scalar coupling constant.  

			If the fermions have masses $m_{1}$ and
$m_{2}$, the total interaction potential including gravity and scalar/vector
interactions can be written in the general form
\begin{equation}
 V(r) = -\frac{Gm_{1}m_{2}}{r}\left(1 + \alpha_{12}\,e^{-r/\lambda}\right),
\label{G+Y potential}
\end{equation}
where
\begin{equation}
	\alpha_{12} \equiv \mp\frac{f_{1}f_{2}}{4\pi Gm_{1}m_{2}}.
\end{equation}
The dimensionless constant $\alpha_{12}$ then characterizes the strength of
the interaction relative to gravity, and its sign depends on type of boson
exchanged.  When $r \ll \lambda$, $|\alpha_{12}| = 1$ indicates a force of
gravitational strength.  		

		Yukawa potentials also arise naturally in models where new gravitational
forces appear from extra spatial dimensions.   String theory requires
there to be more than 3 spatial dimensions, but until recently it
was thought that all the extra dimensions were compactified on the Planck
scale and thus invisible to any conceivable experiment.  However,
much attention recently has focused on a number of string-inspired models in
which all Standard Model particles are confined to the usual 3 spatial
dimensions (a 3-brane) while gravity can ``see''
all dimensions\cite{Antoniadis,Arkani,Sundrum,Shiu,Nussinov}.  In such
models, one would never expect to see the effects of extra dimensions in
Standard Model physics, but their effects would appear in gravitational
physics.  Since so much of the parameter space of gravity remains
unexplored, these effects could have easily escaped detection.  For example,
in models in which the extra dimensions are compact, it is possible that the
compactification radius $r_{c}$ could be as large as
$10^{-3}$ m and thus  would have not been seen in any experiment to date
\cite{Antoniadis,Arkani}.  These models would produce dramatic deviations
from Newtonian gravity at short distances since  they imply that the
$r$-dependence of the gravitational potential between point masses changes
when the particle separation approaches $r_{c}$:
\begin{equation}
V_{\rm grav}(r) = \left\{
        \begin{array}{cl}
      -\displaystyle\frac{G_{4}m_{1}m_{2}}{r} & \mbox{for $r \gg r_{c}$}
\\
        & \\
      -\displaystyle\frac{G_{4 + n}m_{1}m_{2}}{r^{1 + n}} & 
									\mbox{for $r\ll r_{c}$}.
      \end{array}
      \right. 
\end{equation}
Here $n$ is the number of extra spatial dimensions, $G_{4} = G$ is the usual
macroscopic Newtonian gravitational constant, and $G_{4 + n}$ is the more
fundamental gravitational constant for the total $4 + n$ dimensional
spacetime.  Thus, in this model, Newton's law of gravity is merely a
projection of a more fundamental law of gravity onto 3 spatial dimensions,
and the unusual weakness of gravity relative to the other fundamental forces
is attributed to this projection.  One of the striking features of some
recent string models is that compactification can occur over scales much
larger than the Planck scale ($1/ M_{\rm Planck} \sim
10^{-35}$~m)\cite{Antoniadis,Arkani,Nussinov}.  For example, if in a $4 + n$
dimensional spacetime the fundamental mass scale $M_{\rm fund} \sim
M_{\rm EW}$, where $M_{\rm EW}~\sim$~1~TeV is the electroweak scale, then one
expects the compactification scale $r_{c}$ to be given by
\cite{Antoniadis,Arkani,Nussinov}
\begin{equation}
r_{c} \sim \frac{1}{M_{\rm fund}}
         \left(\frac{M_{\rm Planck}}{M_{\rm fund}}\right)^{2/n}
\sim (\mbox{$10^{-19}$ m})(10^{16})^{2/n},
\end{equation}
which yields, 
\begin{equation}
  r_{c} \sim \left\{
     \begin{array}{ll}
       \mbox{$10^{13}$ m}, & \,\,\, n = 1,\\
       \mbox{$10^{-3}$ m}, & \,\,\, n = 2, \\
       \mbox{$10^{-9}$ m}, & \,\,\, n = 3. \\
    \end{array}
   \right.
\end{equation}
 Since no deviations from  Newtonian gravity have been observed for
$r \gtrsim 10^{-3}$~m, theories which suggest that $r_{c}  
\lesssim 10^{-3}$~m  (e.g., $n \geq 2$) are compatible with current experimental
limits.  As one tests gravity over smaller distance scales, the effects of new
extra dimensions  would first appear as  corrections
to the usual gravitational potential. It has been shown that these
corrections for 
$r \gg r_{c}$ have a Yukawa form \cite{Arkani,Kehagias,Floratos}:
\begin{equation}
V_{\rm grav}(r) \sim -\frac{G_{4}m_{1}m_{2}}{r}
                  \left(1 + \alpha_{n} e^{-r/\lambda}\right).
\label{string V}
\end{equation}
Here $\alpha_{n}$ is a composition-independent universal constant that
depends on
$n$ and the nature of the compactification, and the range of the
interaction is $\lambda \sim r_{c}$, the compactification scale.  For
example, for $n$ extra dimensions compactified on an $n$-torus, $\alpha_{n} =
2n$ \cite{Arkani,Kehagias,Floratos}.  It is thus
conceivable that the first evidence supporting the existence of extra spatial
dimensions (and string theory) could come from the detection of a
composition-independent Yukawa modification of Newton's law of gravity.

\subsection{Current Constraints on New Yukawa Forces}

	    Let us now turn to the current laboratory constraints on new Yukawa
forces which arise from a generic potential of the form,
\begin{equation}
	V_{Y}(r) = -\alpha\frac{Gm_{1}m_{2}}{r}e^{-r/\lambda}.
\label{Yukawa V}
\end{equation}
(Here we assume that the interactions are attractive for positive $\alpha$.)
This potential will lead to a violation of the WEP in a gravity experiment if
$\alpha$ is composition-dependent, as in the case of a vector or scalar
interaction.  In addition, even if $\alpha$ is independent of the
composition of the test masses, as in the case of the extra dimension
theories described earlier, small violations of the WEP will still be
present  in an experiment using different materials due to the
``finite-size'' effect \cite{Fischbachbook}.  This effect arises because a
non-uniform Yukawa field will ``capture'' different fractions of two
finite-sized objects having the same mass, but different densities, as will
be the case in the null experiments considered below.   Finally, in addition
to violating the WEP, $V_{Y}(r)$ will also violate the ISL, and so
constraints on $\alpha$ can be inferred from tests of the gravitational ISL.

\begin{figure}
\begin{center}
\includegraphics[scale=.45]{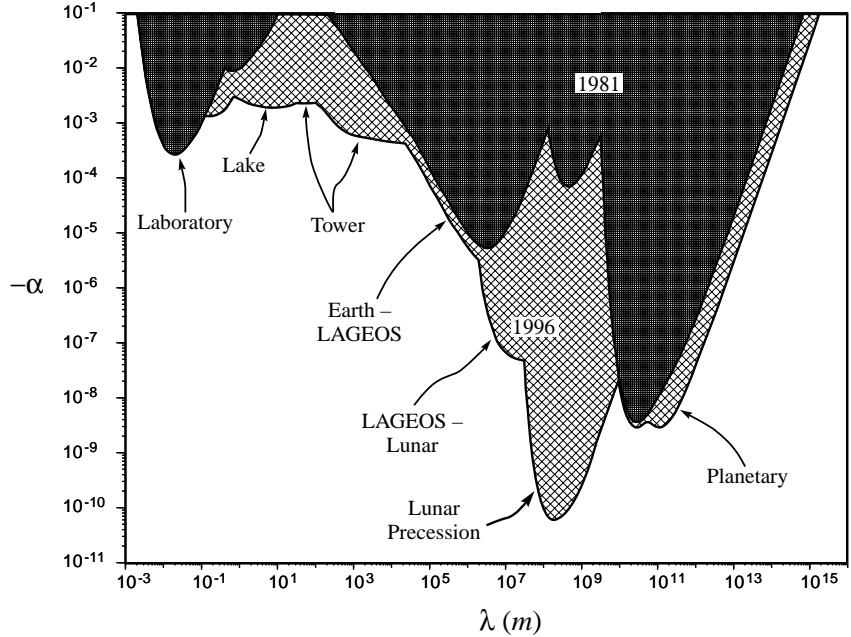}
\end{center}
\caption[]{$2\sigma$ constraints on the coupling constant $\alpha$ as a
function of the range $\lambda$ from composition-independent experiments
\protect\cite{Fischbachbook}.  The dark shaded area indicates the region
excluded as of 1981, and the light hatched region gives the 1996
limits which remain current. }
\label{Yukawa Constraints}
\end{figure}

   As shown in Fig.~\ref{Yukawa Constraints}, the current
experimental constraints on the Yukawa coupling constant $\alpha$ as a
function of range $\lambda$ are quite stringent (allowed $|\alpha|~\ll~1$)
for $10^{-3}$ m $\lesssim \lambda \lesssim 10^{15}$~m, but they fall
exponentially outside this region. Composition-dependent experiments have
set strong limits for specific couplings (e.g., to baryon number) when
$\lambda \gtrsim 10^{15}$~m \cite{Fischbachbook}, but also fall off
exponentially for $\lambda \lesssim 10^{-3}$~m.  

			As discussed in more detail in Ref.~\cite{Long}, current constraints allow
$\alpha \gtrsim 1$ for $\lambda \lesssim 10^{-3}$ m.  For $10^{-4}$ m
$\lesssim \lambda \lesssim 10^{-3}$ m, these limits were obtained from a
test of the gravitational ISL by Mitrofanov and Ponomareva
\cite{Mitrofanov}, but  they still permit a new force
with $\alpha \sim 10^{4}$ for $\lambda \sim 10^{-4}$ m.  However, a new round
of gravity experiments \cite{Long,Carugno,Deufel} should fill in much of this
region of parameter space within the next few years.   At shorter distances,
Casimir/van der Waals forces dominate gravity so the current limits are set
by Casimir force experiments \cite{Long,Lamoreaux,Bordagpreprint,Mohideen}
and are much less restrictive than those obtained from the longer ranged
gravity experiments.  [See also Refs.~\cite{Bordag,MostepanenkoPlates} for
detailed discussions on extracting limits on new forces from Casimir force
experiments.]   The region
$\lambda \lesssim 10^{-4}$~m will remain essentially unexplored until the
next generation of experiments specifically dedicated to search for new
forces is designed and carried out.  We turn next to a discussion of some of
the difficulties likely to be encountered in developing this next round of
experiments.

\section{Problems in Testing Gravity at Very Short Distances}

\subsection{General Problems}

    As noted in the Introduction, a number of authors
\cite{Price,Long,Onofrio,Beane,Giudice} have called attention to the huge gap in
our understanding of gravity at very short distances, and to
its potential to reveal new physics.  The fact that short-distance gravity
experiments can potentially expose the presence of extra spatial dimensions
is particularly tantalizing.  However, since
$|\alpha| \sim 1$ in these string models, the ultimate experimental goal is
quite challenging, namely to set limits $|\alpha| \lesssim 1$ for $\lambda
\lesssim 10^{-3}$ m.  To accomplish this, one has to be able to sense and
distinguish a force of gravitational strength at these distance scales.
Since the current laboratory limits in this region are orders of magnitude
less sensitive than this goal, we discuss briefly some of the
difficulties in studying gravity at short distances.  

		 An obvious problem is scaling \cite{Long}.  Suppose we have two identical
spheres of density $\rho$ and radius $R$.  If we wish to test the law
of gravity between these spheres at short distances, 
the dimensions of the spheres have to be made comparable to the small
distance scales that we are interested in probing.  Since the minimum
separation distance between centers is
$2R$, this leads to a maximum force
\begin{equation}
F_{max} = \frac{G(4\pi\rho R^{3}/3)^{2}}{(2R)^{2}} \propto R^{4}.
\end{equation}
This example illustrates that the gravitational force between
macroscopic objects, which is already quite small, decreases rapidly with
size and separation of the test masses. 

	A second problem in searching for new short-ranged forces is that their
short range limits the effective mass of a body that can participate in the
interaction.  Suppose we have a sphere of uniform density and radius
$R$.  Since gravity is a long-range force,  another identical sphere close
by will interact with all of the mass of the first sphere.  However, if
there exists a new force of range $\lambda \ll R$, then only the layer of
material of thickness $\sim\lambda$ on the surface of the sphere will
interact with external objects, and hence only a fraction $\sim\lambda/R$
of the total mass participates in interactions.  But this problem is actually
much worse in general.  If we have two spheres nearly touching, it is only
the mass within a range $\lambda$ of the contact point that interacts, which
is much less than the fraction $\lambda/R$,  while the gravitational force is
still felt by all the mass.  Therefore, even if the new force is
intrinsically of gravitational strength ($\alpha = 1$) between point masses 
when
$r \ll \lambda$, this new force between macroscopic bodies will usually be
much smaller than the corresponding gravitational force.  This situation was
not encountered in previous longer range gravity experiments  since in those
cases, $\lambda \gtrsim L$, where $L$ was the characteristic size of the test
bodies used.

A third problem occurs when the separation distance is $\lesssim
10^{-4}$ m, where intermolecular forces become
significant.  Since these forces have a power-law form $1/r^{n}$
\cite{Israel}, where
$n$ depends on the geometry of the macroscopic bodies, they grow very
rapidly as $r$ decreases and  overwhelm gravity at very short
distances.  Distinguishing a force of gravitational strength from this
background will be a major challenge.

\subsection{Quantitative Example:  Parallel Plate Gravity Experiment}

\subsection*{Idealized Setup}

			To better appreciate how these problems might arise in actual
experiments, let us now consider a simple experimental setup.  Our
goal here is to estimate the size of the various effects which might appear,
and not to propose the optimal experimental design, and hence we will ignore
practical problems which might be encountered when one actually attempts to
realize such a design. Since we are searching for forces of very short
range, the discussion of the previous section suggests we should have
most, if not all, of the mass of the two test bodies in an experiment
contributing in order to realize the largest possible force.  This means
that we need to have all the mass in one body as close as possible to all the
mass of the second body.  The simplest way to accomplish this is to use 
parallel plates as our test bodies
\cite{MostepanenkoPlates} which maximizes the ``effective mass'' for any
short-range range force.   It then follows that the most appropriate
configuration for searching for new forces between macroscopic bodies is a
parallel-plate experiment analogous to those used to study the Casimir
effect \cite{Casimir}.

\begin{figure}
\begin{center}
\includegraphics[scale=.60, bb=0in 4.5in 8in 10.5in]{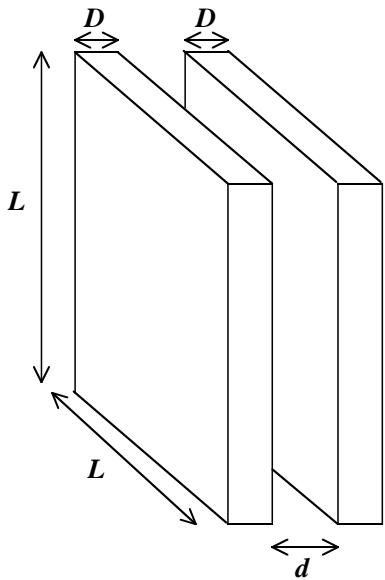}
\end{center}
\caption[]{The idealized parallel plate setup used to quantitatively
estimate the relative magnitudes of the gravitational, Casimir, and Yukawa
forces.  Numerical results were obtained by letting $L = 1$ cm, $D = 1$ mm,
$\rho_{\rm copper} = 8.96 \times 10^{3}$ m$^{3}$, $T = 300$ K, and $d \ll
L$.}
\label{plates}
\end{figure}

		Let us now consider two identical plates of density $\rho$, thickness $D$,
area $A = L^{2}$, separated by a distance $d$ (Figure \ref{plates}).  If we
assume $d \ll L$, we can then safely neglect edge effects and calculate the
pressures between the plates as if $L = \infty$. In addition, we assume
that the plates are perfectly smooth, perfectly conducting, and at
temperature $T = 300$ K.  

\subsection*{Force Formulas}

			We begin our investigation of the
forces between these plates with the known forces, starting with gravity. 
For this particular configuration, the gravitational force acting on the
plates is given by 
\begin{equation}
F_{\rm Gravity}(d) = -2\pi G\rho^{2}L^{2}D^{2},
\label{plate F_G}
\end{equation}
where the minus sign indicates an attractive force.  We see that when the
plates are sufficiently close, the gravitational attraction is
constant, independent of the separation $d$.

			If there are no stray charges, etc., gravity is the dominant force at
large plate separations, but as $d$ decreases, the Casimir force grows
rapidly and quickly overwhelms gravity.  Calculating the Casimir force for
this geometry for real metals can become quite complicated, involving
corrections for finite conductivity and surface roughness
\cite{Lambrecht,CasCorr,LamCor}.  However, for
present purposes we will ignore these difficulties by assuming the plates to
be smooth and perfectly conducting over all frequencies, which should
be a good approximation as long as the plates are not too close. 
However, we will include thermal effects which become large when the plate
separation is large.  The Casimir force between our plates at temperature
$T$ can be written as \cite{Mehra,Brown}
\begin{eqnarray}
   F_{\rm Casimir}(d) & =  & -\frac{\pi^{2}}{240}\frac{\hbar
cL^{2}}{d^{4}}   
    - \frac{\pi kTL^{2}}{d^{3}}\sum_{n = 1}^{\infty}n^{2}\ln\left[
         1 - \exp\left(-\frac{n\pi\hbar c}{kTd}\right)\right]
 \nonumber \\ 
    &  & \mbox{} - \frac{\pi^{2}(kT)^{4}L^{2}}{45(\hbar c)^{3}},
\label{full F_C}
\end{eqnarray}
where $k$ is Boltzmann's constant and factors of $\hbar$ and $c$
have been included for convenience.   Eq.~(\ref{full F_C}) simplifies in the 
two limiting cases \cite{Mehra,Brown}:
\begin{equation}
F_{\rm Casimir}(d) = \left\{
       \begin{array}{cl}
    -1.202\displaystyle\left(\frac{kTL^{2}}{4\pi d^{3}}\right),
&
           \,\,\,\, d \gg \displaystyle\frac{\pi\hbar c}{kT}, \\
             & \\
     -\displaystyle\frac{\pi^{2}\hbar cL^{2}}{240 d^{4}}, &  
           \,\,\,\, d \ll \displaystyle\frac{\pi\hbar c}{kT}.
\end{array}\right. 
\label{limited F_C}
\end{equation}

   Having obtained expressions for the known forces acting between the
plates in this idealized setup, let us now
determine the forces arising from possible new interactions.  The attractive
Yukawa potential between point masses as given by Eq.~(\ref{Yukawa
V}) leads to a force between the plates (with $d \ll L$) given by
\begin{equation}
F_{\rm Yukawa}(d) = -2\pi\alpha\lambda^{2}G\rho^{2}L^{2}
       \left(1 - e^{-D/\lambda}\right)^{2}e^{-d/\lambda}.
\label{F_Y}
\end{equation}
We then notice that the ratio of this Yukawa force to the gravitational force
in Eq.~(\ref{plate F_G}) is
\begin{equation}
\frac{F_{\rm Yukawa}(d)}{F_{\rm Gravity}(d)} =
\alpha\left(\frac{\lambda}{D}\right)^{2}
    \left(1 - e^{-D/\lambda}\right)^{2}e^{-d/\lambda}.
\end{equation}
If $\lambda \ll D$, then
\begin{equation}
\frac{F_{\rm Yukawa}(d)}{F_{\rm Gravity}(d)} \simeq
\alpha\left(\frac{\lambda}{D}\right)^{2}e^{-d/\lambda}.
\end{equation}
Thus, even if the Yukawa coupling is intrinsically of gravitational
strength ($\alpha =1$), the actual Yukawa force is suppressed relative to
gravity not only by the usual exponential factor
$e^{-d/\lambda}$, but also by $(\lambda/D)^{2}$ which arises because only a
fraction $\lambda/D$ of the total mass of each plate contributes to the
Yukawa force.  This effect was discussed earlier and clearly illustrates how
a short-ranged force intrinsically of gravitational strength is strongly
suppressed in an experiment using macroscopic bodies.

\subsection*{Numerical Results}

	Having found the general formulas for all the forces that we will be
considering, let us now obtain numerical values
for the  following setup.   We assume that the plates have dimensions
$L \times L \times D$, where $L = 1$ cm and $D = 1$ mm, which are roughly
comparable to the values used in some of the current short distance gravity
experiments \cite{Long,Carugno}.  Since our previous calculations assumed
$d \ll L$, we focus our attention on the region
$10^{-8}$ m $\lesssim d \lesssim 10^{-3}$~m.  Next we will assume that the
plates are made of pure copper which has a density $\rho = 8.96 \times
10^{3}$ kg/m$^{3}$.  Except for the new force parameters
$\alpha$ and $\lambda$, our problem is now completely specified.

			Using these numbers, we  first calculate the known forces, gravity and
Casimir, for the plates.  As discussed earlier, the gravitational force
under the conditions assumed here is constant and given by Eq.~(\ref{plate
F_G}).  Substituting the parameters given above yields
\begin{equation}
F_{\rm Gravity} = 3.37 \times 10^{-12}\,\,\,\mbox{Newtons}.
\label{plate gravity number}
\end{equation}

\begin{figure}[t]
\begin{center}
\includegraphics[scale=.80, bb=.5in 5.5in 9in 10.5in]{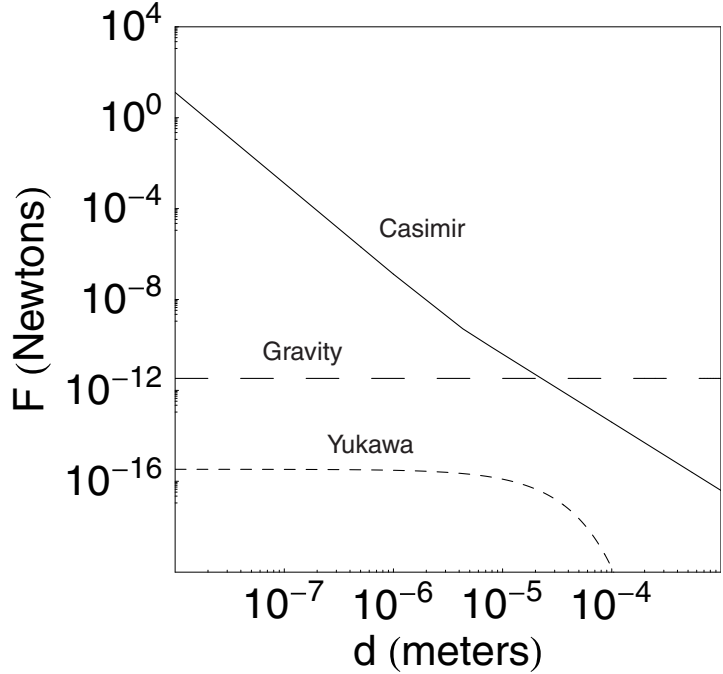}
\end{center}
\caption[]{The Casimir, gravitational, and Yukawa ($\alpha = 1$, $\lambda =
10^{-5}$ m) forces between the parallel plates shown in
Fig.~\protect\ref{plates}.}
\label{parallel plate forces figure}
\end{figure}

		To determine the Casimir force for this configuration, we  use 
Eq.~(\ref{full F_C}).  The cross-over distance
$d_{c}$, where temperature-dependent effects become important at $T = 300$ K,
is
\begin{equation}
   d_{c} = \frac{\pi\hbar c}{kT} = 2.4 \times 10^{-5}\,\,\mbox{m} = 24
\,\,\mu{\rm m}.
\end{equation}
Graphs of the Casimir force using Eq.~(\ref{limited F_C}), and the
gravitational force between the plates, are shown together in Figure
\ref{parallel plate forces figure}, and numerical values of these forces at
various distances can be found in Table \ref{forces table}.  The
gravitational and Casimir forces are equal to each other when $T = 300$ K at
$d = 2.3 \times 10^{-5}$ m = 23 $\mu$m, which just happens to coincide with
$d_{c}$ here.  Thus, for $d \lesssim 23$ $\mu$m, the Casimir
force will dominate gravity in this setup.
\begin{table}[t]
\caption{The magnitudes of the gravitational, Casimir, and Yukawa
(using $\alpha = 1$ and $\lambda = 10^{-5}$~m) forces arising in the
idealized parallel plate experiment discussed in the text.  Here
$F_{\rm Back} = F_{\rm Gravity} + F_{\rm Casimir}$ is the total background
force against which the signal of $F_{\rm Yukawa}$ must be seen.}
\begin{center}
\begin{tabular}{ccccc} \hline
$d$ (m) \hspace{.25in}& \hspace{.25in} $F_{\rm gravity}$ (N)\hspace{.25in}
&
\hspace{.25in}$F_{\rm Casimir}$ (N) \hspace{.25in}&
\hspace{.25in} $F_{\rm Yukawa}$ (N) 
&
\hspace{.25in} $F_{\rm Yukawa}/F_{\rm Back}$ \\  
\hline\noalign{\smallskip} 
$10^{-3}$ & $3.4 \times 10^{-12}$ & $4.2 \times 10^{-17}$ & $1.3 \times
10^{-59}$ & $4 \times 10^{-48}$ \\
$10^{-4}$ & $3.4 \times 10^{-12}$ & $4.0 \times 10^{-14}$ & $1.5 \times
10^{-20}$ & $4 \times 10^{-9}$ \\
$10^{-5}$ & $3.4 \times 10^{-12}$ & $4.0 \times 10^{-11}$ & $1.2 \times
10^{-16}$ & $3 \times 10^{-6}$ \\
$10^{-6}$ & $3.4 \times 10^{-12}$ & $1.3 \times 10^{-7}$ & $3.0 \times
10^{-16}$ & $2 \times 10^{-9}$ \\
$10^{-7}$ & $3.4 \times 10^{-12}$ & $1.3 \times 10^{-3}$ & $3.3 \times
10^{-16}$ & $3 \times 10^{-13}$ \\
$10^{-8}$ & $3.4 \times 10^{-12}$ & $1.3 \times 10^{1}$ & $3.4 \times
10^{-16}$ & $3 \times 10^{-17}$ \\ \hline
\end{tabular}
\end{center}
\label{forces table}
\end{table}

\begin{figure}[t]
\begin{center}
\includegraphics[scale=.6,bb=0in 4in 8in 10.5in]{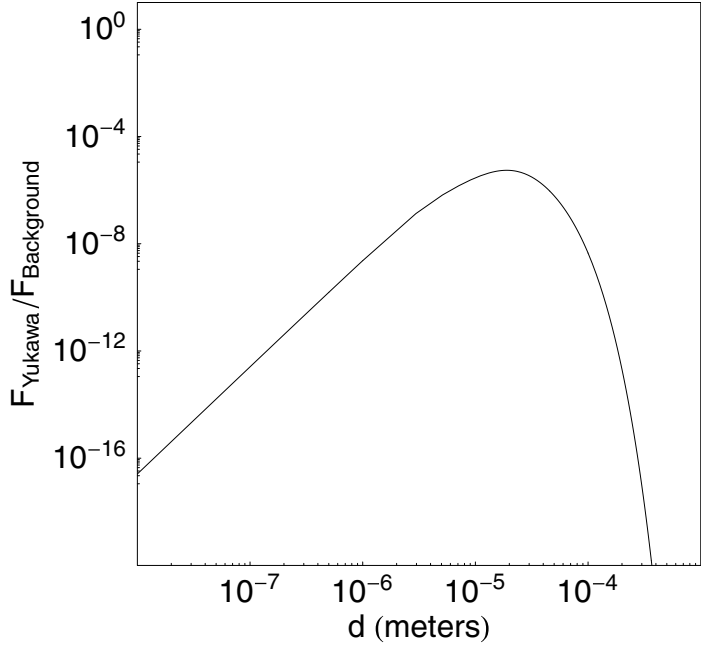}
\end{center}
\caption[]{The ratio of the Yukawa ($\alpha = 1$, $\lambda =
10^{-5}$ m) force to the background force $F_{\rm Back} = F_{\rm Gravity} +
F_{\rm Casimir}$ for the parallel plate setup shown in
Fig.~\protect\ref{plates}.}
\label{force ratio figure}
\end{figure}

			Now let us turn to new Yukawa forces,  which are characterized by 
two free parameters, the relative strength $\alpha$ and the range $\lambda$. 
If for illustrative purposes we consider a force of gravitational strength
($\alpha = 1$) and set
$\lambda = 10^{-5}$ m, then Eq.~(\ref{F_Y}) yields $F_{\rm Yukawa}(d)$
exhibited in Figure~\ref{parallel plate forces figure} and
Table~\ref{forces table}.  We see that when
$d
\ll
\lambda$, the force becomes constant:
\begin{equation}
F_{\rm Yukawa}(d\ll \lambda) = 3.37 \times 10^{-16}\,\,\mbox{Newtons},
\end{equation}
which is $(\lambda/D)^{2} = 10^{-4}$ times smaller than the corresponding
gravitational force given by Eq.~(\ref{plate gravity number}).  Thus, as
explained earlier, even though the Yukawa force  between {\em point}
particles is of gravitational strength at short distances, $F_{\rm Yukawa}$
is much smaller than gravity for these macroscopic plates.  We also clearly
see that
$F_{\rm Yukawa}/F_{\rm Back}$ is maximized when $d\sim \lambda$ and falls
off rapidly from this plate separation (Fig.~\ref{force ratio figure}).  This
is because $F_{\rm Yukawa}$ levels off when $d \lesssim \lambda$ while
$F_{\rm Casimir}$ continues to increase via a power-law ($1/d^{4}$ if $d
\lesssim d_{c}$).

\subsection*{Constraining New Short-Ranged Yukawa Forces}

	This analysis using an obviously idealized setup reveals the two critical
problems that will be encountered in devising  experiments using macroscopic
bodies to search for very short-ranged Yukawa interactions of gravitational
strength ($\alpha
\sim 1$).  The first is that the absolute magnitude of such a force will be
very small, possibly even smaller than the gravitational force if $\alpha$ is
small.  Thus, an experiment must be sensitive to the smallest possible
forces.  Second, since the Casimir background force grows rapidly as the
separation decreases, one must be able to extract the signal of a very weak
force from a background of very strong intermolecular forces.   A direct
attack on this problem would be to attempt to calculate as accurately as
possible the background forces in a gravity experiment, and to then subtract
these from the observed force to set limits using what remains
\cite{Long,Bordagpreprint,Bordag,MostepanenkoPlates}.  However,
recent experiments studying the Casimir force reveal the difficulty
of accurately calculating the background forces to better than 1\%
\cite{Lamoreaux,Mohideen,Lambrecht,CasCorr,LamCor,LamCom,LamandMoh}. 
While this approach is still possible, we will describe in the next section a new method
of performing a null short-distance gravity experiment specifically designed to directly
subtract out the unwanted background effects.  Calculating intermolecular forces precisely
then becomes much less important.

\section{Very Short Distance Null Gravity Experiments}

		Some of the best constraints on Yukawa interactions come from tests of the
WEP \cite{Fischbachbook}.  In these experiments, one compares the
accelerations of compositionally-different test bodies toward a common source
body.  Any differences in these accelerations can then be attributed to
non-gravitational forces.  We wish to utilize the same principle in a
short-distance gravity experiment, and thus avail ourselves to the extreme
sensitivity of such experiments.

\subsection{Null Experiment \#1}

\begin{figure}[t]
\begin{center}
\includegraphics[scale=.55, bb=0in 2in 8in 10.5in]{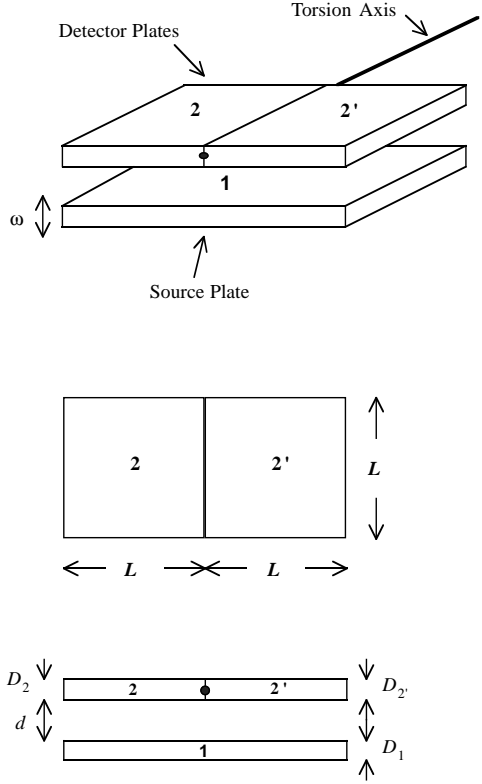}
\end{center}
\caption[]{Schematic design for Very Short Distance Null Experiment \#1. 
See text for details.}
\label{Experiment 1}
\end{figure}

Inspired by two ongoing short distance experiments \cite{Long,Carugno}, a
possible design for one such experiment is shown in Figure
\ref{Experiment 1}.  It consists of two parallel plates, a source plate 1
and a detector plate made of two smaller plates 2 and $2'$.  The source
plate is driven sinusoidally with angular frequency $\omega$ such that the
separation distance $d$ is given by
\begin{equation}
d(t) = d_{0} + d_{1}\cos\omega t.
\end{equation}
Instead of detecting a force, this experiment would be sensitive to a
torque, modulated by the frequency $\omega$, about an axis passing along
the boundary where plates 2 and $2'$ are joined, as shown in
Fig.~\ref{Experiment 1}.  If we assume that the plates are conducting, the
net torque $\tau_{\rm net}$ on the detector plate will arise from
contributions from the gravitational, Casimir, and possibly, Yukawa forces:
\begin{eqnarray}
\tau_{\rm net} & = & \tau_{\rm Gravity} + \tau_{\rm Casimir} + \tau_{\rm
Yukawa} \nonumber \\
 & = & \frac{L}{2}\left[ (F_{2}^{\rm Gravity} - F_{2'}^{\rm Gravity})
      + (F_{2}^{\rm Casimir} - F_{2'}^{\rm Casimir}) \right.
       \nonumber \\
      &  & \mbox{} \left. +
     (F_{2}^{\rm Yukawa} - F_{2'}^{\rm Yukawa}) \right],
\end{eqnarray}
where $F_{i}^{\rm Gravity}$, $F_{i}^{\rm Casimir}$, and $F_{i}^{\rm Yukawa}$
are the gravitational, Casimir, and Yukawa forces on plates $i = 2$ and $2'$
respectively.  One then selects plates 2 and $2'$ such that the torque
$\tau_{\rm Gravity} + \tau_{\rm Casimir}$ due to background forces vanishes
while $\tau_{\rm Yukawa} \neq 0$ if $\alpha_{1i} \neq 0$. [Here we allow
for the possibility that $\alpha_{12}$ and $\alpha_{12'}$, the Yukawa
couplings between the materials comprising plates 1 and 2, and 1 and $2'$
respectively, are different.]

  At very small separations, $F_{i}^{\rm Gravity}$ will be negligible (and
independent of $\omega$ to first approximation) , but it is still easy to
make $\tau_{\rm Gravity}$ vanish anyway.  Using Eq.~(\ref{plate F_G}), we see
that
\begin{equation}
|F_{2}^{\rm Gravity} - F_{2'}^{\rm Gravity}| = 2\pi GL^{2}\rho_{1}D_{1}
\left|\rho_{2}D_{2} - \rho_{2'}D_{2'}\right|,
\end{equation}
where $\rho_{i}$ and $D_{i}$ are the density and plate thickness of the
$i$th plate.  Then 
\begin{equation}
(\rho_{2}D_{2} = \rho_{2'}D_{2'}) \Rightarrow (M_{2} = M_{2'}) \Rightarrow
\tau_{\rm Gravity} = 0, 
\end{equation}
so if the detector plates 2 and $2'$ have the same mass, the
gravitational torque will vanish.

  Of course, the much bigger challenge is choose materials for plates 2 and
$2'$ such that the Casimir torque $\tau_{\rm Casimir}$ also vanishes.  If the
plates were perfectly conducting, this would be the case since
Eq.~(\ref{full F_C}) would be identical for all such plates with the same
surface area.  However, the finite conductivity of real metallic plates
becomes very important when the plate separation $d \sim \lambda_{P}$,
where $\lambda_{P} = 2\pi c/\omega_{P}$, and $\omega_{P}$ is the plasma
frequency of the metal.  Still, it was shown recently
\cite{Lambrecht} that  the Casimir force
between pairs of copper and gold plates are equal to a good approximation for
separations $d \gtrsim 10^{-6}$ m at $T = 0$.  Such calculations
are difficult for real materials, but this certainly raises the hope that it
is possible to choose appropriate plates 2 and $2'$ such that
\begin{equation}
F_{2}^{\rm Casimir} - F_{2'}^{\rm Casimir} 
\simeq 0 \,\,\, \Rightarrow \,\,\, \tau_{\rm
Casimir} \simeq 0.
\label{F_C cancellation}
\end{equation}
At the very least, one should be able to fabricate the plates using two
different isotopes  of the same element (e.g., $^{24}$Mg and $^{26}$Mg) such
that Eq.~(\ref{F_C cancellation}) is satisfied.  The underlying premise of
the ``iso-electronic'' effect (IE)  is that to a good approximation the
Casimir effect depends on the {\em electronic} properties of the materials,
and hence is largely independent of their {\em nuclear} properties.  By
contrast, the gravitational interaction, and virtually all proposed new
Yukawa interactions, involve couplings to both electrons and nucleons. 
Hence, subtracting out the electronic contributions by choosing two isotopes
of some material, or by choosing materials with similar electronic
properties (such as Cu and Au), we can enhance the signal from a new Yukawa
force while simultaneously reducing the Casimir background.

The remaining torque after the gravitational and Casimir torques have
been suppressed might be due to a new Yukawa force.  Using Eq.~(\ref{F_Y}),
the net torque due to a putative Yukawa force would be 
\begin{eqnarray}
\tau_{\rm Yukawa} & = & 2\pi G\lambda^{2}(L/2)L^{2}\rho_{1}
      \left(1 - e^{-D_{1}/\lambda}\right)e^{-d/\lambda} \nonumber \\
&  & \times\left[\alpha_{12}\rho_{2}\left(1 - e^{-D_{2}/\lambda}\right)
       - \alpha_{12'}\rho_{2'}\left(1 - e^{-D_{2'}/\lambda}\right)\right].
\label{full tau_Y}
\end{eqnarray}
If $\lambda \ll D_{i}$, then Eq.~(\ref{full tau_Y}) simplifies to
\begin{equation}
\tau_{\rm Yukawa}  \simeq  2\pi
G\lambda^{2}(L/2)L^{2}\rho_{1}e^{-d/\lambda}
\left[\alpha_{12}\rho_{2}
       - \alpha_{12'}\rho_{2'}\right].
\label{simplified tau_Y}
\end{equation}
If the Yukawa force arises from an extra-dimensional modification of
Newtonian gravity such that $\alpha_{12} = \alpha_{12'} = \alpha_{n}$,
Eq.~(\ref{simplified tau_Y}) reduces to
\begin{equation}
\tau_{\rm Yukawa}  \simeq 
2\pi\alpha_{n} G\lambda^{2}(L/2)L^{2}\rho_{1}\rho_{2}e^{-d/\lambda}
       \left(1 - \frac{\rho_{2'}}{\rho_{2}}\right).
\end{equation}
This result depends on the  difference in the mass densities $\rho_{2} -
\rho_{2'}$ for a simple reason:  When $\lambda \ll D_{i}$, the force only
sees the mass within a distance $\lambda$ of the surface.  Thus, if plates 2
and
$2'$ have different densities, the effective mass seen by the Yukawa force
will be different and a net torque arises due to the ``finite-size''
effect discussed earlier.  It is then clear that one should choose
materials 2 and $2'$ such that
$\rho_{2}$ and $\rho_{2'}$ differ as much as possible while still ensuring
that the Casimir torque vanishes.  For the gold/copper and
$^{24}$Mg/$^{26}$Mg combinations suggested above, 
\begin{equation}
1 - \frac{\rho_{2'}}{\rho_{2}} \simeq \left\{
     \begin{array}{cl}
     0.32 & \,\,\,\,\, \mbox{2 = Au, $2'$ = Cu}, \\
        &  \\
     0.077 & \,\,\,\,\, \mbox{2 = $^{26}$Mg, $2'$ = $^{24}$Mg}.
     \end{array}
     \right.
\label{tau suppression}
\end{equation}
The hope is that the suppression factor Eq.~(\ref{tau suppression}) is more
than compensated by the reduction of the unwanted background torques.

\subsection{Null Experiment \#2}

\begin{figure}[t]
\begin{center}
\includegraphics[scale=.5,bb=0in 5in 8in 11in]{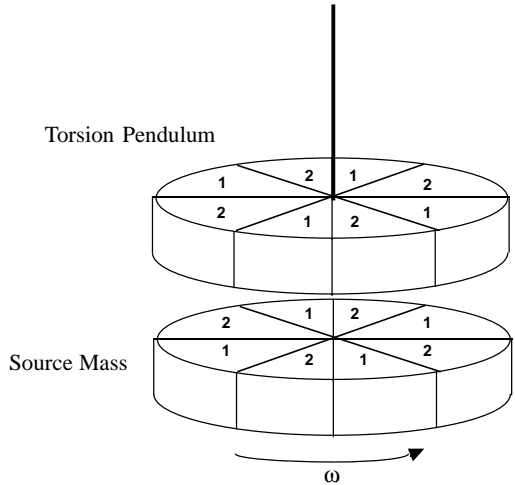}
\end{center}
\caption[]{Schematic design for Very Short Distance Null Experiment
\#2.  See text for details.}
\label{Experiment 2}
\end{figure}

We conclude this section by briefly describing another possible design for a
short distance null experiment motivated by another set of
gravity experiments \cite{Deufel,Boynton}.  As shown schematically in
Figure~\ref{Experiment 2}, this experiment consists of two disks, one
serving as the source mass, while the other (detector mass) is the
pendant of a torsion pendulum.  Each disk is divided into
alternating wedges made of two different materials 1 and 2.  The
wedges and materials 1 and 2 are designed such that no Casimir torque would
arise when the source disk rotates below the pendulum.  For example, as
indicated above, 1 and 2 might be gold/copper or $^{24}$Mg/$^{26}$Mg which
would significantly reduce the Casimir torque.  Then, if the separation
between the disks is small, the remaining torque on the pendulum
would arise from a putative Yukawa force because the effective mass within a
distance of $\lambda$ will be different for alternating wedges.  

\section{Discussion}

	 It is clear that there is significant motivation for testing Newtonian
gravity at very short distances.  However, as we have seen, new problems
will face experimentalists who wish to extend current constraints on
the Yukawa coupling constant $\alpha$ down to ranges $\lambda \lesssim
10^{-4}$ m. Our aim in this paper has been to point out some of the most
obvious difficulties, but there may be others that have passed unnoticed. 
While we have not addressed the significant issue of improving the
sensitivity of experiments to very small forces, we have taken the first
steps towards dealing with the problem of detecting small forces against a
large intermolecular force background.  Our schematic designs for null
experiments are meant as illustrations of the principles involved in
canceling background forces (the ``iso-electronic'' and ``finite-size''
effects), and hence are not intended to suggest optimal designs.  It is
hoped that experimentalists, who face the realities of imperfect materials
and incomplete theories,  can extract some useful ideas  from these schematic
designs or, perhaps, will be able to point to flaws which preclude them from
working as actual experiments.  Finally, we conclude with the encouraging
note that since so little is known about gravity at separations
$\lesssim 10^{-4}$ m, virtually {\em any} good experiment in this region
will tell us something new.
   
\subsection*{Acknowledgments}

We wish to thank P. Boynton, G. Carugno, C. Deufel, D. Koltick, A.
Lambrecht, J. Mullen, R. Newman,  R. Reifenberger, S. Reynaud, and C.
Talmadge for very useful discussions.  D. Krause also acknowledges the
support of Wabash College and Purdue University, and this work was supported
in part by the U.S. Department of Energy under Contract No. DE-AC
02-76ER01428.

\end{document}